\documentclass[english]{article}
\usepackage[T1]{fontenc}
\usepackage[latin9]{inputenc}
\usepackage{amsmath}
\usepackage{setspace}
\makeatletter
\newenvironment{lyxlist}[1]
{\begin{list}{}
{\settowidth{\labelwidth}{#1}
 \setlength{\leftmargin}{\labelwidth}
 \addtolength{\leftmargin}{\labelsep}
 }}
{\end{list}}

\makeatother

\usepackage{babel}

\begin{document}

\title{Comment on {}``Security analysis and improvements of arbitrated
quantum signature schemes'' }

\author{Tzonelih Hwang%
\thanks{Corresponding Author%
}, Yi-Ping Luo, Song-Kong Chong}
\maketitle
\begin{abstract}
Recently, Zou et al. {[}Phys. Rev. A 82, 042325 (2010){]} demonstrated
that two arbitrated quantum signature (AQS) schemes are not secure,
because an arbitrator cannot arbitrate the dispute between two users
when a receiver repudiates the integrity of a signature. By using
a public board, Zou et al. proposed two AQS schemes to solve the problem.
This work shows that the same security problem may exist in Zou et
al.'s schemes. Moreover, a malicious verifier, Bob, can actively negate
a signed order if he wants to. This attack, a special case of denial-of-service
(DoS) attack mentioned in {[}Phys. Rev. Lett. 91, 109801 (2003){]},
is important in quantum cryptography. Bob may get some benefits with
this DoS attack, since he can actively deny Alice's signed order without
being detected. This work also shows that a malicious signer can reveal
the verifier's secret key without being detected by using Trojan-horse
attacks.

\end{abstract}

\section{Introduction}

The quantum signature, which provides the authenticity and non-repudiation
of quantum states on an insecure quantum channel \cite{Zen2002,Cho2011_3},
is one of the most important topics of research in quantum cryptography.
The quantum signature can provide unconditional security by exploiting
the principles of quantum mechanics, such as the no-cloning theory
and measurement uncertainty. Two basic properties are required in
a quantum signature \cite{Zen2002} :
\begin{enumerate}
\item Unforgeability: Neither the signature verifier nor an attacker can
forge a signature or change the content of a signature. The signature
should not be reproducible by any other person.
\item Undeniability: A signatory, Alice, who has sent the signature to the
verifier, Bob, cannot later deny having provided a signature. Moreover,
the verifier Bob cannot deny the receipt of a signature. 
\end{enumerate}
The first quantum signature was proposed by Gottesman and Chuang \cite{Got2001}.
Subsequently, a variety of quantum signature schemes have been proposed
\cite{Zen2002,Cho2011_3,Yan2010,Cho2011,Cur2008,Zen2008,Li2009,Yan2010-1,Wan2005_3,Lee2004,Wan2006_3,Wen2006,Wen2007_2}.
Zeng et al. \cite{Zen2002} proposed an arbitrated quantum signature
(AQS) scheme based on the correlation between Green-Horne-Zeilinger
(GHZ) states and quantum one-time pads. However, Curty et al. \cite{Cur2008}
pointed out that this AQS scheme \cite{Zen2002} is not clearly described
and that the security statements claimed by the authors are incorrect.
In response \cite{Zen2008}, Zeng provided a more detailed presentation
and proof to Zeng et al.'s original AQS scheme \cite{Zen2002}. To
improve the transmission efficiency and to reduce the implementation
complexity of \cite{Zen2002,Zen2008}, Li et al. \cite{Li2009} proposed
an AQS scheme using Bell states and claimed that their improvements
can preserve the merits in the original scheme \cite{Zen2002,Zen2008}. 

In an AQS scheme, the arbitrator plays a crucial role. When a dispute
arises between users, the arbitrator should be able to arbitrate the
dispute. In other words, the arbitrator should be able to solve a
dispute when a verifier, Bob, repudiates the receipt of a signature
or, in particular, when the verifier repudiates the integrity of a
signature, i.e., Bob admits receiving a signature but denies the correctness
of the signature. The latter dispute implies one of the following
three cases \cite{Zou2010}:
\begin{description}
\item [{\textmd{(1)}}] Bob told a lie;
\item [{\textmd{(2)}}] The signatory Alice sent incorrect information to
Bob;
\item [{\textmd{(3)}}] An eavesdropper Eve disturbed the communications.
\end{description}
As the arbitrator in \cite{Zen2002,Zen2008,Li2009} cannot solve the
dispute when Bob claims that the verification of a signature is not
successful, Zou et al. \cite{Zou2010} considered that these schemes
are not valid AQS schemes because the security requirement of a quantum
signature, i.e., undeniability, is not satisfied. 

By using a public board, Zou et al. also proposed two AQS schemes
to solve the problem. However, this study demonstrates that the same
security problem may exist in Zou et al.'s schemes. In their schemes,
when Bob announces that the verification of a signature is not successful,
the arbitrator may not be able to arbitrate the dispute mentioned
above. Moreover, a malicious verifier, Bob, can actively negate a
signature if he wants to. This attack, a special case of denial-of-service
(DoS) attack mentioned in \cite{Cai2003}, is important in quantum
cryptography. Bob may get some benefits with this DoS attack, since
he can actively deny Alice's signature without being detected. In
addition, this study attempts to demonstrate that a malicious signer,
Alice, can reveal Bob's secret key without being detected by using
Trojan-horse attacks \cite{Cai2006,Deng2005_3}.

The rest of this paper is organized as follows. Section 2 reviews
one of Zou et al.'s schemes. Section 3 discusses the problems with
the scheme. Finally, Section 4 summarizes the result.

\section{Review of Zou et al.'s first signature scheme}

Zou et al.'s first AQS scheme \cite{Zou2010} is briefly explained
in the following scenario. Alice, the message signatory, wants to
sign a quantum message $\left|P\right\rangle $ to a signature verifier,
Bob, via the assistance of an arbitrator, Trent. Suppose that Alice
and Bob share a secret key $K\in\left\{ 0,1\right\} ^{*}$ and that
the quantum message to be signed is $\left|P\right\rangle =\left|P_{1}\right\rangle \otimes\left|P_{2}\right\rangle \otimes...\otimes\left|P_{n}\right\rangle $,
where $\left|K\right|\geq2n$, $\left|P_{i}\right\rangle =\alpha_{i}\left|0\right\rangle +\beta_{i}\left|1\right\rangle $,
and $1\leq i\leq n$. In order to protect the quantum message, the
quantum one-time-pad encryption $E_{K}$ \cite{Boy2003} and the unitary
transformation $M_{K}$ used in the schemes are defined as follows.

\begin{equation}
E_{K}\left(\left|P\right\rangle \right)=\overset{n}{\underset{i=1}{\bigotimes}}\sigma_{x}^{K_{2i-1}}\sigma_{z}^{K_{2i}}\left|P_{i}\right\rangle ,\end{equation}

\begin{equation}
M_{K}\left(\left|P\right\rangle \right)=\overset{n}{\underset{i=1}{\bigotimes}}\sigma_{x}^{K_{i}}\sigma_{z}^{K_{i\oplus1}}\left|P_{i}\right\rangle ,\end{equation}

\noindent where $\left|P_{i}\right\rangle $ and $K_{i}$ denote the
$i$th bit of $\left|P\right\rangle $ and $K$, respectively, and
$\sigma_{x}$ and $\sigma_{z}$ are the respective Pauli matrices.

To prevent the integrity of a signature from being repudiated by Bob,
Zou et al. proposed two AQS schemes: the AQS scheme using Bell states
and the AQS without using entangled states. In this paper, we only
review their AQS scheme using Bell states.

Suppose that Alice wants to sign an $n$-qubit quantum message $\left|P\right\rangle $
to Bob. In order to perform the signature, three copies of $\left|P\right\rangle $
are necessary. The scheme proceeds as follows:
\begin{lyxlist}{00.00.0000}
\item [{\textbf{Initialization}}] \noindent \textbf{phase:}\end{lyxlist}
\begin{description}
\item [{Step$\;\mathbf{\mathit{I}}\mathbf{1}.$}] The arbitrator Trent
shares the secret keys $K_{A}$ and $K_{B}$ with Alice and Bob, respectively,
through some unconditionally secure quantum key distribution protocols.
\item [{Step$\; I\mathbf{2}.$}] Alice generates $n$ Bell states, $\left|\psi_{i}\right\rangle =\frac{1}{\sqrt{2}}\left(\left|00\right\rangle _{AB}+\left|11\right\rangle _{AB}\right)$,
where $1\leq i\leq n$; the subscripts $A$ and $B$ denote the first
and second particles of the Bell state, respectively. After that,
Alice sends all $B$ particles to Bob in a secure and authenticated
way \cite{Cur2002,Bar2002}.\end{description}
\begin{lyxlist}{00.00.0000}
\item [{\textbf{Signing$\;\:$phase:}}]~\end{lyxlist}
\begin{description}
\item [{Step$\; S\mathbf{1}.$}] Alice chooses a random number $r\in\left\{ 00,01,10,11\right\} ^{n}$
to encrypt all $\left|P\right\rangle $'s, i.e., $\left|P'\right\rangle =E_{r}\left(\left|P\right\rangle \right)$.
\item [{Step$\; S\mathbf{2}.$}] Alice generates $\left|S_{A}\right\rangle =E_{K_{A}}\left(\left|P'\right\rangle \right)$.
\item [{Step$\; S\mathbf{3}.$}] Alice combines each $\left|P_{i}'\right\rangle $
with the first particle $A$ of each Bell state. Then, each original
Bell state becomes a three-particle entangled state, \textbf{\footnotesize \[
\left|\phi_{i}\right\rangle _{PAB}=\left|P_{i}'\right\rangle \otimes\left|\psi_{i}\right\rangle _{AB}=\frac{1}{2}\left[\begin{array}{c}
\left|\Phi_{PA}^{+}\right\rangle _{i}\left(\alpha_{i}^{'}\left|0\right\rangle +\beta_{i}^{'}\left|1\right\rangle \right)_{B}+\left|\Phi_{PA}^{-}\right\rangle _{i}\left(\alpha_{i}^{'}\left|0\right\rangle -\beta_{i}^{'}\left|1\right\rangle \right)_{B}+\\
\left|\Psi_{PA}^{+}\right\rangle _{i}\left(\alpha_{i}^{'}\left|1\right\rangle +\beta_{i}^{'}\left|0\right\rangle \right)_{B}+\left|\Psi_{PA}^{-}\right\rangle _{i}\left(\alpha_{i}^{'}\left|1\right\rangle -\beta_{i}^{'}\left|0\right\rangle \right)_{B}\end{array}\right],\]
}where $\left|\Phi_{PA}^{+}\right\rangle ,\left|\Phi_{PA}^{-}\right\rangle ,\left|\Psi_{PA}^{+}\right\rangle $,
and $\left|\Psi_{PA}^{-}\right\rangle $ are the four Bell states
\cite{Kwi1995}.
\item [{Step$\; S\mathbf{4}.$}] Alice performs a Bell measurement on each
pair $\left|\phi_{i}\right\rangle _{PA}$ and obtains the measurement
results $\left|M_{A}\right\rangle =\left(\left|M_{A}^{1}\right\rangle ,\left|M_{A}^{2}\right\rangle ,\ldots,\left|M_{A}^{n}\right\rangle \right)$,
where $\left|M_{A}^{i}\right\rangle \in\left\{ \left|\Phi_{PA}^{+}\right\rangle _{i},\left|\Phi_{PA}^{-}\right\rangle _{i},\left|\Psi_{PA}^{+}\right\rangle _{i},\left|\Psi_{PA}^{-}\right\rangle _{i}\right\} $,
and $1\leq i\leq n$.
\item [{Step$\; S\mathbf{5}.$}] Alice sends $\left|S\right\rangle =\left(\left|P'\right\rangle ,\left|S_{A}\right\rangle ,\left|M_{A}\right\rangle \right)$
to Bob.\end{description}
\begin{lyxlist}{00.00.0000}
\item [{\textbf{Verification}}] \textbf{phase:}\end{lyxlist}
\begin{description}
\item [{Step$\;\mathbf{\mathit{V}1}.$}] Bob encrypts $\left|P'\right\rangle $
and $\left|S_{A}\right\rangle $ with $K_{B}$ and sends the quantum
ciphertext $\left|Y_{B}\right\rangle =E_{K_{B}}\left(\left|P'\right\rangle ,\left|S_{A}\right\rangle \right)$
to Trent.
\item [{Step$\;\mathbf{\mathit{V}2}.$}] Trent decrypts $\left|Y_{B}\right\rangle $
with $K_{B}$ and obtains $\left|P'\right\rangle $ and $\left|S_{A}\right\rangle $.
Next, he encrypts $\left|P'\right\rangle $ with $K_{A}$ and obtains
$\left|S_{T}\right\rangle $. If $\left|S_{T}\right\rangle =\left|S_{A}\right\rangle $
\cite{Li2009,Buh2001}, then Trent sets the verification parameter
$V=1$; otherwise, he sets $V=0.$
\item [{Step$\; V\mathbf{3}.$}] Trent recovers $\left|P'\right\rangle $
from $\left|S_{T}\right\rangle $. Then, he encrypts $\left|P'\right\rangle ,\left|S_{A}\right\rangle $,
and $V$ with $K_{B}$ and sends the quantum ciphertext $\left|Y_{T}\right\rangle =E_{K_{B}}\left(\left|P'\right\rangle ,\left|S_{A}\right\rangle ,V\right)$
to Bob.
\item [{Step$\; V\mathbf{4}.$}] Bob decrypts $\left|Y_{T}\right\rangle $
and obtains $\left|P'\right\rangle ,\left|S_{A}\right\rangle $, and
$V$. If $V=0$, Bob rejects the signature; otherwise, Bob continues
to the next step.
\item [{Step$\; V\mathbf{5}.$}] Based on Alice's measurement results $M_{A}$,
Bob can obtain $\left|P'_{B}\right\rangle $ from the $B$ particles
received from the Step $I\mathbf{2}$, according to the principle
of teleportation \cite{Li2009}. Next, he compares $\left|P_{B}'\right\rangle $
with $\left|P'\right\rangle $. If $\left|P_{B}'\right\rangle =\left|P'\right\rangle $,
Bob informs Alice to publish $r$ and proceeds to the next step; otherwise,
he rejects the signature.
\item [{Step$\; V\mathbf{6}.$}] Alice publishes $r$ on the public board.
\item [{Step$\; V\mathbf{7.}$}] Bob recovers $\left|P\right\rangle $
from $\left|P'\right\rangle $ by $r$ and holds $\left(\left|S_{A}\right\rangle ,r\right)$
as Alice's signature for the quantum message $\left|P\right\rangle $.
\end{description}

\section{Discussion on Zou et al.'s scheme}

This section discusses problems that could arise in Zou et al.'s scheme
if precautions are not taken. We first present a DoS attack by using
undeniability dilemma and give an example to show that a verifier
can actively negate a signature without being detected to get some
benefits in his favor. Then, we introduce Trojan-horse attacks to
Zou et al.'s scheme.

\subsection{Undeniability dilemma - A Denial-of-service (DoS) attack}

In Zou et al.'s scheme, the signatory Alice uses a random number $r$
to protect the quantum message $\left|P\right\rangle $ (i.e., $\left|P'\right\rangle =E_{r}\left(\left|P\right\rangle \right)$)
before signing it. After the verification by the arbitrator Trent,
Bob recovers $\left|P'_{B}\right\rangle $ and compares it with $\left|P'\right\rangle $.
Once Bob informs Alice that $\left|P_{B}'\right\rangle =\left|P'\right\rangle $,
Alice publishes $r$ on the public board, which is assumed to be free
from being blocked, injected, or altered. Finally, Bob recovers $\left|P\right\rangle $
from $\left|P'\right\rangle $ by $r$ and retains $\left(\left|S_{A}\right\rangle ,r\right)$
as Alice's signature. 

It appears that if Bob informs Alice to publish $r$ on the public
board, then he cannot disavow the integrity of the signature. In accordance
with this logic, Zou et al. considered that the use of the public
board can prevent the denial attack from Bob. However, if Bob claims
that $\left|P_{B}'\right\rangle \neq\left|P'\right\rangle $ in Step
$V5$ before requesting the value of $r$ from Alice, then Trent cannot
arbitrate the dispute between Alice and Bob because one of the following
three possible cases may occur.
\begin{enumerate}
\item Bob told a lie: In this case, Bob decides to forego the recovery of
the message $\left|P\right\rangle $ due to some reasons;
\item Alice sent incorrect information to Bob: In Step $S3$, Alice deliberately
generated $\left|\phi_{i}\right\rangle $ using another message $\left|\hat{P}_{i}'\right\rangle $
with $\left|\hat{P}_{i}'\right\rangle \neq\left|P_{i}'\right\rangle $
or generated $\left|S\right\rangle =\left(\left|P'\right\rangle ,\left|S_{A}\right\rangle ,\left|M'_{A}\right\rangle \right)$
with $\left|M'_{A}\right\rangle \neq\left|M_{A}\right\rangle $ in
Step $S5$;
\item Eve disturbed the communication.
\end{enumerate}
Apparently, when Bob claims that $\left|P_{B}'\right\rangle \neq\left|P'\right\rangle $
in this case, Trent cannot solve the dispute. Hence, Bob can perform
the DoS attack by negating the signature from Alice without being
detected. Furthermore, as also pointed out in \cite{Zou2010}, Alice
is able to publish an arbitrary value $r'\left(\neq r\right)$ such
that the oringinal signature cannot be verified successfully by Bob,
which is also contradictory to the undeniable requirement of a signature
scheme. 

This problem could be serious if the signature occurs in an electronic
order system, where Alice is a buyer and Bob, a company. Bob is able
to negate a signed order from Alice if the current market situation
is not in his favor. In such a case, it does not matter whether Bob
can obtain the value $r$ to recover the signed order from Alice,
because Bob knows that due to the order, he will lose a fortune.  Similarly,
by controlling the value of $r$, Alice is also able to select a situation
favorable to her for completing the signature process. 

The same dilemma may occur in Zou et al.'s second AQS scheme.

\subsection{Trojan-horse attacks}

In Zou et al.'s scheme, there are two transmissions of the same quantum
signals, first from Alice to Bob and then from Bob to Trent. Therefore,
the malicious Alice can reveal Bob's secret key without being detected
by using Trojan-horse attacks \cite{Cai2006,Deng2005_3}. As pointed
out in \cite{Cho2011}, there are two ways to use Trojan-horse attacks:
invisible photon eavesdropping (IPE)\cite{Cai2006} and delay photon
eavesdropping \cite{Deng2005_3}. Here, we discuss the IPE attack
on Zou et al.'s scheme and demonstrate that Alice can obtain Bob's
secret key without being detected. It should be noted that Alice can
also use the delay photon eavesdropping to reveal Bob's secret key
in the same way.

In order to reveal Bob's secret key $K_{B}$, Alice can use the IPE
attack on the communications in Step $S5$ and Step $V1$ as follows:
\begin{description}
\item [{Step$\; S\mathbf{5a}.$}] Alice first prepares a set of eavesdropping
states, $D^{i}\in\left\{ \frac{1}{\sqrt{2}}\left(\left|00\right\rangle +\right.\right.$
$\left.\left.\left|11\right\rangle \right)_{d_{1}^{i}d_{2}^{i}}\right\} $,
as invisible photons, where the subscripts $d_{1}^{i}$ and $d_{2}^{i}$
represent the first and second photons, respectively, in $D^{i}$,
$1\leq i\leq n$. For each state in $\left|P'\right\rangle $ (or
$\left|S_{A}\right\rangle $), Alice inserts $d_{1}^{i}$ as an invisible
photon to that state and forms a new sequence $\left|P'\right\rangle ^{d_{1}}$
($\left|S_{A}\right\rangle ^{d_{1}}$). Next, Alice sends $\left|S\right\rangle ^{d_{1}}=\left(\left|P'\right\rangle ^{d_{1}},\left|S_{A}\right\rangle ,\left|M_{A}\right\rangle \right)$
to Bob.
\item [{Step$\;\mathbf{\mathit{V}1a}.$}] Bob encrypts $\left|P'\right\rangle ^{d_{1}}$
and $\left|S_{A}\right\rangle $ with $K_{B}$ and sends the quantum
ciphertext $\left|Y_{B}\right\rangle ^{d_{1'}}=E_{K_{B}}(\left|P'\right\rangle ^{d_{1}},$
$\left|S_{A}\right\rangle )$ to Trent. Before Trent receives the
quantum ciphertext $\left|Y_{B}\right\rangle ^{d{}_{1'}}$, Alice
captures $d_{1'}$ from $\left|Y_{B}\right\rangle ^{d_{1'}}$ and
measures $d_{1'}d_{2}$ together with the Bell measurement. According
to the measurement result of $d_{1'}^{i}d_{2}^{i}$, Alice can obtain
Bob's secret key $K_{B}^{2i-1,2i}$. 
\end{description}
It should be noted that Alice can similarly use the process mentioned
above to obtain Bob's secret key $K_{BT}$ in Zou et al.'s second
AQS scheme. Since both schemes are susceptible to Trojan-horse attacks,
Bob can deny having verified a signature. 

To prevent the scheme from Trojan-horse attacks, it is well-know that
two additional devices, a wavelength filter and a photon number splitter
(PNS) can be added to the protocol. By letting the received photons
pass through both devices, the photons with different wavelength or
the delay photons will not exist or will be detected \cite{Gis2002,Deng2005_3}.

\section{Conclusions}

This paper has pointed out security flaws in Zou et al.'s AQS schemes,
in which Trent cannot arbitrate a dispute between Alice and Bob when
Bob claims a failure in the signature verification phase. Besides,
a malicious verifier, Bob, can actively negate a signed order from
Alice without being detected to get some benefits in his favor. In
addition, we demonstrate that a malicious signatory can reveal the
verifier's secret key by launching Trojan-horse attacks on Zou et
al.'s AQS scheme. How to design an AQS scheme without the DoS attack
and how to construct an AQS scheme free from Trojan-horse attacks
without using any hardware device will be an interesting future research.

\section*{Acknowledgment}

The authors would like to thank the anonymous reviewers for their
very valuable comments that enhance the clarity of this paper a lot.
We also like to thank the National Science Council of Republic of
China and the Research Center of Quantum Communication and Security,
National Cheng Kung University, Taiwan, R.O.C. for financial support
of this research under Contract No. NSC 100-2221-E-006-152-MY3 and
D100-36002, respectively.

\bibliographystyle{IEEEtran}
\addcontentsline{toc}{section}{\refname}\bibliography{YIPING,ISLAB}

\end{document}